\documentclass[letterpaper, 10 pt, conference]{ieeeconf}  % Comment this line out if you need a4paper
\pdfoutput=1 
\usepackage{ntheorem}

\makeatletter
\newcounter{subhyp} 
\let\savedc@hyp\c@hyp

\newcommand{\normhyp}{%
  \let\c@hyp\savedc@hyp % revert to the old one
  \renewcommand\thehyp{\arabic{hyp}}%
} 
\makeatother
\usepackage{fixltx2e}

\usepackage[cmex10]{amsmath}
\usepackage{fainekos-macros}
\usepackage{wrapfig,floatflt,yfonts}
\usepackage{mathtools}
\usepackage{colonequals}
\usepackage{enumerate}
\usepackage{graphicx}
\usepackage{subcaption}
\graphicspath{ {figures/} }
\DeclareGraphicsExtensions{.pdf}
\usepackage{multirow}
\usepackage{todonotes}
\usepackage{algpseudocode}
\usepackage{textcomp}
\usepackage{color}
\usepackage{url}
\usepackage{dsfont}
\usepackage{stmaryrd} 
\usepackage{todonotes}
\usepackage{listings}
\usepackage{colonequals}
\usepackage{xspace}
\usepackage{booktabs}
\usepackage{array}
\usepackage{afterpage}
\usepackage{rotating}
\usepackage{pdflscape}
\usepackage{tikz}
\usetikzlibrary{arrows}
\usetikzlibrary{positioning}

\usepackage{textcomp}

\usepackage{ifthen}
\newboolean{TECHREP}
\setboolean{TECHREP}{true}

\usepackage{algorithm}
\usepackage{algpseudocode}
\IEEEoverridecommandlockouts                              % This command is only needed if 
\title{\LARGE \bf
\textsc{ViSpec}: A graphical tool for elicitation of MTL requirements
}

\author{Bardh Hoxha$^{1}$, Nikolaos Mavridis$^{2}$ and Georgios Fainekos$^{1}$% <-this % stops a space
\thanks{$^{1}$ Bardh Hoxha and Georgios Fainekos are with the School of Computing, Informatics and Decision Systems Engineering, Arizona State University
        {\tt\small \{bhoxha,fainekos\}@asu.edu}}%
\thanks{$^{2}$Nikolaos Mavridis is with the Interactive Robots and Media Lab and NCSR Demokritos {\tt\small nmav@alum.mit.edu}}%
}

\begin{document}

\newcommand{\ViSpec}{\text{\sc ViSpec}}
\newcommand{\vspacemod}[1]{ \ifthenelse{\boolean{TECHREP}}{\vspace{0cm}}{\vspace{#1}}}

\maketitle
\thispagestyle{empty}
\pagestyle{empty}

%%%%%%%%%%%%%%%%%%%%%%%%%%%%%%%%%%%%%%%%%%%%%%%%%%%%%%%%%%%%%%%%%%%%%%%%%%%%%%%%
\begin{abstract}

One of the main barriers preventing widespread use of formal methods is the elicitation of formal specifications. 
Formal specifications facilitate the testing and verification process for safety critical robotic systems. 
However, handling the intricacies of formal languages is difficult and requires a high level of expertise in formal logics that many system developers do not have. 
In this work, we present a graphical tool designed for the development and visualization of formal specifications by people that do not have training in formal logic. 
The tool enables users to develop specifications using a graphical formalism which is then automatically translated to Metric Temporal Logic (MTL). 
In order to evaluate the effectiveness of our tool, we have also designed and conducted a usability study with cohorts from the academic student community and industry. 
Our results indicate that both groups were able to define formal requirements with high levels of accuracy. 
Finally, we present applications of our tool for defining specifications for operation of robotic surgery and autonomous quadcopter safe operation. 

\end{abstract}

%%%%%%%%%%%%%%%%%%%%%%%%%%%%%%%%%%%%%%%%%%%%%%%%%%%%%%%%%%%%%%%%%%%%%%%%%%%%%%%%
%!TEX root = root.tex
\pdfoutput=1 
\section{Introduction}

As robots become commercially available, their correct operation is of paramount importance. Especially for safety critical systems, safety must be guaranteed. As for example in autonomous vehicles \cite{WongpiromsarnEtAl2012tecs} and medical robots \cite{muradore2011robotic,DBLP:conf/hybrid/KouskoulasRPK13}. 

%\item Why is the definition of natural language specifications a problem? (eventually folk vs eventually formal, superficial similarity (phonology), OR (folk) vs OR (formal) vs XOR(formal), gricean maxims, conversational implicature(pragmatics), “there are 4 objects on the table” -> “there are exactly 4 objects on the table”).

Safety requirements are usually expressed in natural language, which is inherently ambiguous, in general. When it is used for defining system specifications, this ambiguity may lead to misunderstandings between development teams that may result in increased costs and delays in development. If the misunderstandings are not detected, then a product that does not meet the intended specifications will be developed. 

Ideally, specifications should be defined in a mathematical language, using formal logics. This not only removes ambiguity, but also allows system developers to utilize a vast set of methods \cite{TripakisD09model} that have been developed by the academic community for testing and verification of systems. The academic community has also developed automatic tools such as \staliro \cite{AnnapureddyLFS11tacas,hoxhatowards},  \textsc{Fapas} \cite{Yordanov2013261}, SpaceEx \cite{FrehseCAV11}, CheckMate \cite{SilvaK00acc}, \textsc{Flow} \cite{Chen+Abraham+Sankaranarayanan-2013-Flow}, Breach \cite{donze10cav}, C2E2 \cite{duggirala2013verification}, KeYmaera \cite{PlatzerQ08ijcar} and \textsc{Strong} \cite{deng2013strong} that enable developers to conduct system testing and verification. 

%On the other hand, formal specifications can also be used for high level planning for time critical robot missions \cite{lignos2015provably,smith2011optimal,KaramanS08cdc,BhatiaKV10icra,WongpiromsarnTM10hscc,kim2014revision,guo2013revising}.

Even though it has been shown, that utilizing formal specifications can lead to improved testing and verification \cite{FainekosSUY12acc}, the industry still utilizes natural language as the premier approach in defining specifications. One may conjecture that the most important reason for doing so is because the development of specifications through a formal logic requires a level of mathematical training that many users may not have \cite{vinter1998applying}. Furthermore, even for expert users, writing formal specifications is an error prone task \cite{holzmann2002logic}. As a result, the industry has been less willing to utilize formal specifications in their processes.

\ifthenelse{\boolean{TECHREP}}{
\begin{figure*}[th]
\vspacemod{-10pt} % in text
  \centering
  \includegraphics[width=13cm]{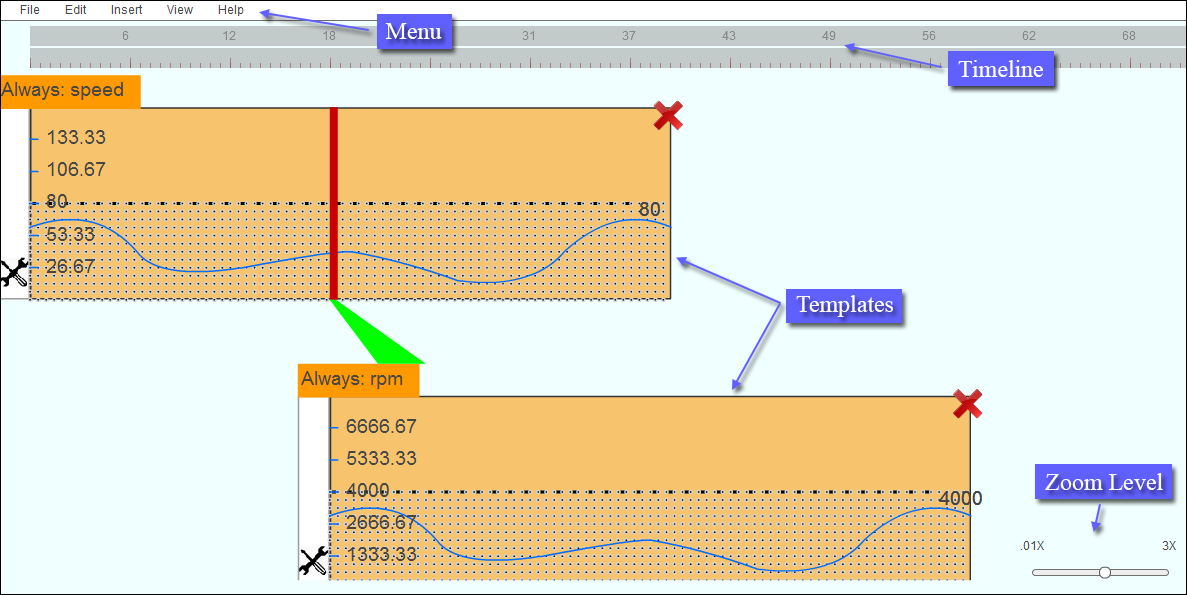}
  \caption{Overview of the graphical user interface of the MTL specification tool. The example shown represents the MTL specification $\phi = \Box_{[0,40]}( (speed<80) \rightarrow \Box_{[0,40]} (rpm < 4000) )$.}
  \label{fig:overviewMTL}%
  \vspacemod{-20pt} % in text
\end{figure*}
}{}
In this work, we present a graphical formalism that enables non-expert users to develop formal specifications for control systems. The formalism enables the visualization of a large fragment of MTL. The main challenge in the development of the formalism lies in finding the right balance between expressive power and ease-of-use. It is designed for use with systems and signals and enables both event and time based specifications. This is the first time that a visual formal language representation is developed for specifications for Cyber-Physical Systems (CPS). Here by CPS we define any system that has discontinuous nonlinear dynamics and complex safety critical requirements. Prime examples are medical robotics and autonomous vehicles. A specification visualization tool has been developed based on the graphical formalism presented in this work. To evaluate the usefulness of the tool in terms of usability and ease-of-use, we have conducted a usability study.

\noindent \textsc{Summary of Contributions:}
\begin{itemize}
\item We present a graphical formalism that enables the development of formal specifications. 
\item We present the visual specification tool based on the graphical formalism. 
\item We conducted a usability study to evaluate the tool. 
\item Through the usability study we proved that both non-expert users and expert users are able to define formal requirements accurately using the tool, and derived suggestions for improvement of the tool.
\item We present applications of the tool for real-world robots.
\end{itemize}

\noindent \textsc{Related works:}
In order to help address the formal specification challenge, various graphical formalisms have been studied in the past \cite{smith2001events, alfonso2004visual, kugler2005temporal,autili2007graphical,zhang2010timed, srinivas2013graphical}. The most relevant works appear in \cite{autili2007graphical} and \cite{zhang2010timed}. In \cite{autili2007graphical}, the authors extend Message Sequence Charts and UML 2.0 Interaction Sequence Diagrams to propose a scenario based formalism called Property Sequence Chart (PSC). The formalism is mainly developed for specifications on concurrent systems. In \cite{zhang2010timed}, PSC is extended to Timed PSC which enables the addition of timing constructs to specifications.

In terms of usability studies for formal requirements very few works exist. In \cite{vinter1998applying}, the authors study the ability of expert users to develop requirements in Z. A related usability study for requirement representation is presented in \cite{lignos2015provably}, where the authors  present and evaluate a system for generating, troubleshooting and executing controllers for robots using natural language.

%\todo{Nikolaos:mention robotics at least two more times }

%!TEX root = root.tex
\pdfoutput=1 
\section{Visual Specification Tool}
\label{sec:ViSpecTool}

The Visual Specification Tool (\ViSpec{})\footnote{Available at \url{https://sites.google.com/a/asu.edu/s-taliro/vispec}} enables the development of formal specifications for CPS. Users can develop requirements in a graphical formalism which is then translated to Metric Temporal Logic (MTL) \cite{Koymans90}. %The MTL formula can then be used for both for high level planning \cite{smith2011optimal,KaramanS08cdc,BhatiaKV10icra,WongpiromsarnTM10hscc} or  testing and verification of systems \cite{donze10cav,AnnapureddyLFS11tacas}. 

The topic of capturing requirements through graphical formalisms has been studied in the past \cite{smith2001events, alfonso2004visual, kugler2005temporal, autili2007graphical, zhang2010timed}. However, to the best of the authors’ knowledge, the work presented here is the first attempt to do so specifically aimed for the development of specifications for CPS. The initial idea for the graphical formalism was first presented in \cite{hoxhatowards} while the tool was still in the early stages of development. However, in this work we present an updated version of the tool along with its usability study. The improvements over the previous version include: a more streamlined interface; an updated representetion of signals in the interface; and an updated template definition process.

For CPS specifications, it is often needed to account for both timing and event sequence occurrence. Both of these are necessary for reasoning over systems and signals. Consider the specification $\Box_{[0,5]}((speed > 100) \rightarrow \Box_{[0,5]}(rpm > 4000))$. It states that whenever within the first 5 seconds, the \textit{vehicle speed} goes over 100, then from that moment on, the \textit{engine speed (rpm)}, for the next 5 seconds, should always be over 4000. Here both the sequence and timing of the events are of critical importance.

%\cite{smith2001events, zanolin2003approach,alfonso2004visual,braberman2005scenario, kugler2005temporal, autili2007graphical, zhang2010timed}.

To ensure that the tool can be utilized by non-expert users, the following goals for the tool are defined: 1) The user interface is intuitive to use, i.e, it does no have a high learning curve; 2) The visual representation of the requirements is visualy distinct and unambiguous; 3) There is a one-to-one mapping from the visual representation of the requirement and the corresponding requirement in MTL.
%Heuristic Rule 1 - The user interface is easy to use, i.e does no have a high learning curve.
%Heuristic Rule 2 - The visual representation of the requirements is clear and unambiguous.
%Heuristic Rule 3 - There is a one-to-one mapping from the visual representation of the requirement and the corresponding requirement in Metric Temporal Logic. 

\begin{table*}[th]
  \centering
  \caption{Classes of specifications expressible with the graphical formalism}
    \begin{tabular}{p{3cm}p{13.6cm}}
    \toprule
    Specification Class & Explanation \\
    \midrule
    Safety & Specifications of the form $\Box \phi$ used to define specifications where $\phi$ should always be true. \\
    Reachability & Specifications of the form $\Diamond \phi$ used to define specifications where $\phi$ should be true at least once in the future (on now). \\
    Stabilization & Specifications of the form $\Diamond \Box \phi$ used to define specifications that, at least once, $\phi$ should be true and from that point on, stay true. \\
    Recurrence & Specifications of the form $\Box \Diamond \phi$ used to define specifications that, it is always the case, that at some point in the future, $\phi$ is true.  \\
    Implication & Specifications of the form $\phi \rightarrow \psi$ requires the $\psi$ should hold when $\phi$ is true. \\
    Reactive Response & Specifications of the form $N(\phi \rightarrow M\psi)$, where $N$ and $M$ are temporal operators, used to define an implicative response between two specifications where the timing of $M$ is relative to timing of $N$. \\
    Conjunction & Specifications of the form $\phi \wedge \psi$ used to define the  conjunction of two sub-specifications \\
    Non-strict Sequencing & Specifications of the form $N(\phi \wedge M\psi)$, where $N$ and $M$ are temporal operators, used to define a conjunction between two specifications where the timing of $M$ is relative to timing of $N$. \\
    \bottomrule
    \end{tabular}%
  \label{tab:specClasses}%
  \vspacemod{-0.4cm}
\end{table*}%

The set of specifications that can be generated from this graphical formalism is a proper subset of the set of MTL specifications. Formally, the following grammar produces the set of formulas that can be expressed by the proposed graphical formalism:

\begin{minipage}{\linewidth}
\begin{lstlisting}[mathescape]
S $\longrightarrow$ $\neg$T | T
T $\longrightarrow$ A | B | C
A $\longrightarrow$ P | (P$\wedge$A) | (P$\Rightarrow$A) 
B $\longrightarrow$ $\Box_\Ic$D | $\Diamond_\Ic$D
C $\longrightarrow$ $\Box_\Ic\Diamond_\Ic$D | $\Diamond_\Ic\Box_\Ic$D
D $\longrightarrow$ $p$ | ($p$$\rightarrow$A) | ($p$$\wedge$A) | ($p$$\rightarrow$B) | ($ p $$\wedge$B)
P $\longrightarrow$ $p$ | $\Box_\Ic p$ | $\Diamond_\Ic p$
\end{lstlisting}
\end{minipage}
where $p$ is an atomic proposition. In practice, the atomic propositions are automatically derived from the templates.

Throughout the development process of the formalism, it was noticed that the more expressive the formalism, the more challenging to use it became. Therefore, we focused on several widely used classes of specifications which are described in Table \ref{tab:specClasses}. Examples of the classes of specifications are presented in the rest of this section. 

To make the tool easier to use, we placed several constraints on the types of signals used. Specifically, the signals and requirements are one dimensional. This enables clear and structured visualization on a two dimensional user interface. 

\ifthenelse{\boolean{TECHREP}}{
In Fig. \ref{fig:overviewMTL}, the user interface of the tool is presented along with its most critical components. The user interface is composed of a menu, horizontal timeline, rectangular blocks called templates, and a zoom scroll. While the passage of time is represented horizontally, the sequence of events is presented vertically. The formulas are generated from templates as well as the connections between them.}{} 

The main building blocks of the formalism are templates. These are used for defining temporal logic operators, their timing intervals, and the expected signal shape. The user starts with an empty template and a setup assistant presents the user with a sequence of dialog boxes that aid in the development of the template. The process is context dependent where each option selection leads to a potentially different set of options for the next step. 

The first step in the template definition process is to define the temporal operator. Among the choices (and their corresponding MTL symbols) are: \textit{Always} ($\Box$), \textit{At Least Once} ($\Diamond$), \textit{Eventually Always} ($\Diamond \Box$), \textit{Repeatedly Often and Finally} ($\Box \Diamond$), and \textit{now}. The options available enable users to define a wide range of specifications. The following sections will present examples of a subset of formulas that can be generated using this graphical formalism. 

After the temporal operator is selected, the user sets the timing bounds for it. Many users might have difficulty defining timing bounds, especially for specifications with temporal operators such as \textit{Eventually Always} ($\Diamond \Box$) and \textit{Repeatedly Often and Finally} ($\Box\Diamond$). To illustrate the process, the tool provides a fill-in-the-blanks sentence format to the user. For example, if the operator \textit{Eventually Always}  is selected, the user will have to complete the following sentence with the timing bounds: ``Eventually, between \underline{\hspace{0.6cm}} and \underline{\hspace{0.6cm}} seconds, the signal will become true, and from that point on, will stay true in the next \underline{\hspace{0.6cm}} to \underline{\hspace{0.6cm}} seconds". The set of timing intervals are visualized with color shaded regions in the template. 

\begin{figure}[h]
\vspacemod{-10pt} % in text
% \vspacemod{-10pt} % on top of page
\begin{center}
\includegraphics[width=6.5cm]{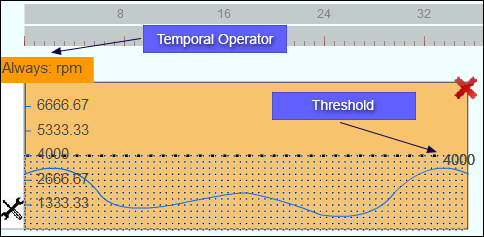}
\end{center}
\vspacemod{-5pt}
\caption{Example \ref{exmp:safety}: The graphical formalism for the \textit{Safety} MTL specification $\phi_{1}=\Box_{[0,36]} (rpm < 4000)$.}
\label{fig:rpmL4000}
\vspacemod{-10pt} % in text
\end{figure}

The next step in the process is in defining whether the predicate will evaluate to true when the signal is above or below a set threshold. For example, for the \textit{Always} ($\Box$) operator, a signal is selected that is either always above or below a specified threshold. Once either option is selected, various signals that fit the requirement are automatically generated and presented visually. Instead of drawing the signal, the user will select from one of the generated options. Consider the following example:

\begin{exmp}
A specification from the fragment of MTL formulas called \textit{Safety} MTL specifications is presented. Specifically, the specification $\phi_{1}=\Box_{[0,36]} (rpm < 4000)$. The formula states that in the next 36 seconds, \textit{engine speed} should always be less than 4000. The corresponding graphical formalism for this formula is presented in Fig. \ref{fig:rpmL4000}. Note that, in regards to the specification, the signal can be of any shape as long as it is always below the 4000 threshold.
\label{exmp:safety}
\end{exmp}

\begin{figure}[h]
\vspacemod{-5pt} % in text
% \vspacemod{-10pt} % on top of page
\begin{center}
\includegraphics[width=6.5cm]{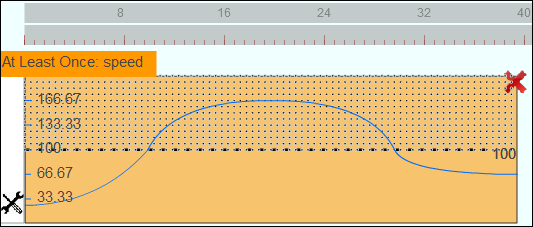}
\end{center}
\vspacemod{-5pt}
\caption{Example \ref{exmp:reachability}: The graphical formalism for the \textit{Reachability} MTL specification $\phi_{2}=\Diamond_{[0,39]} (speed > 100)$.}
\label{fig:speedG100}
\vspacemod{-10pt} % in text
\end{figure}

Consider the following example for the \textit{At Least Once} ($\Diamond$) temporal operator:

\begin{exmp}
A specification from the fragment of MTL formulas called \textit{Reachability} MTL specifications is presented. Specifically, the specification $\phi_{2}=\Diamond_{[0,39]} (speed > 100)$. The formula states that eventually, within the next 39 seconds, the vehicle speed will go over 100. The corresponding graphical formalism for this formula is presented in Fig. \ref{fig:speedG100}. Again, in regards to the specification, the signal can be of any shape as long as at one point, within the timing bounds of the temporal operator, it is above the 100 threshold. 
\label{exmp:reachability}
\end{exmp}

\ifthenelse{\boolean{TECHREP}}{

\begin{figure}[h]
\vspacemod{-5pt} % in text
% \vspacemod{-10pt} % on top of page
\begin{center}
\includegraphics[width=6.5cm]{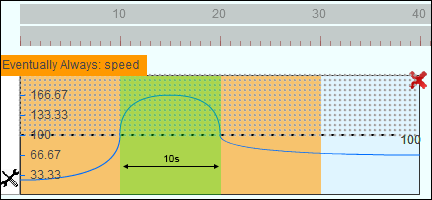}
\end{center}
\vspacemod{-5pt}
\caption{Example \ref{exmp:evenAlways}: The graphical formalism for the MTL specification $\phi_{3}=\Diamond_{[0,30]}\Box_{[0,10]} (speed > 100)$.}
\label{fig:evenAlways}
% \vspacemod{-10pt} % in text
\vspacemod{-5pt} % in text
\end{figure}

For the \textit{Eventually Always} ($\Diamond\Box$) operator, at least once in the timing interval of the eventually operator, the signal should go above the threshold and stay there for the entire timing interval of the always operator. Two types of shading will indicate the timing bounds of the MTL operators. 

\begin{exmp}
Consider the specification $\phi_{3}=\Diamond_{[0,30]}\Box_{[0,10]} (speed > 100)$. The formula states that at some point in the first 30 seconds, the vehicle speed will go over 100 and stay above for 10 seconds. The corresponding graphical formalism for this formula is presented in Fig. \ref{fig:evenAlways}.
\label{exmp:evenAlways}
\end{exmp}

For the \textit{Repeatedly Often and Finally} ($\Box\Diamond$) operator, an oscillating signal is presented where two types of shading indicate the timing intervals for each MTL operator. Consider the following example: 

\begin{exmp}
The specification $\phi_{4}=\Box_{[0,30]}\Diamond_{[0,10]} (speed > 100)$ is presented. The formula states that at every timestep of the simulation in the first 30 seconds, the speed will go over 100 within the next 10 seconds. The corresponding graphical formalism for this formula is presented in Fig. \ref{fig:alwaysEven}. No matter how far to the left or right the green shaded region is moved, contained within the orange region, there is always a point where the signal is above the threshold. Recall that the signal is automatically generated so that it satisfies the options previously selected.
\label{exmp:alwaysEven}
\end{exmp}

\begin{figure}[h]
\vspacemod{-5pt} % in text
% \vspacemod{-10pt} % on top of page
\begin{center}
\includegraphics[width=6.5cm]{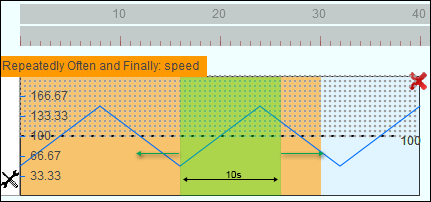}
\end{center}
\vspacemod{-5pt}
\caption{Example \ref{exmp:alwaysEven}: The graphical formalism for the MTL specification $\phi_{4}=\Box_{[0,30]}\Diamond_{[0,10]} (speed > 100)$.}
\label{fig:alwaysEven}
\vspacemod{-5pt} % in text
\end{figure}

The next important concept in this graphical formalism is the relationship between templates. 

First, the sequence relationship between two templates is presented. Assume that the first template is already created. If another template is added below it, then an order in the execution of the events is defined. The second template is only considered if the first template is evaluated to true. Formally, there is an implication relationship from the first template to the second. Consider the following example:

\begin{exmp}
The specification $\phi_{5}=(\Diamond_{[0,40]}(speed > 100)) \rightarrow (\Diamond_{[0,30]} (rpm >3000)) $ is presented. The formula states that if, within 40 seconds, the vehicle speed is above 100 then within 30 seconds from time 0, the engine speed should be over 3000. The corresponding graphical formalism for this formula is presented in Fig. \ref{fig:speedImpRpm}.
\label{exmp:speedImpRpm}
\end{exmp}

\begin{figure}[h]
\vspacemod{-5pt} % in text
% \vspacemod{-10pt} % on top of page
\begin{center}
\includegraphics[width=6.5cm]{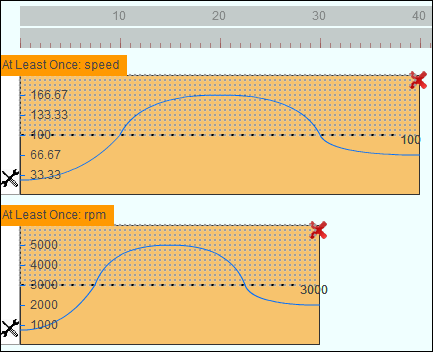}
\end{center}
\vspacemod{-5pt}
\caption{Example \ref{exmp:speedImpRpm}: The graphical formalism for the MTL specification $\phi_{5}=(\Diamond_{[0,40]}(speed > 100)) \rightarrow (\Diamond_{[0,30]} (rpm >3000)) $.}
\label{fig:speedImpRpm}
\vspacemod{-5pt} % in text
\end{figure}

A second type of relationship enables the user to establish conjunction between two events. To achieve this, templates can be grouped. This is indicated by a bold black box. Doing so requires that both templates evaluate to true. Consider the following example:

\begin{exmp}
Specification $\phi_{6}=(\Box_{[0,40]}(speed < 100)) \wedge (\Box_{[0,40]} (rpm < 4000)) $. The formula states that, within 40 seconds, the vehicle speed should be less than 100 and the engine speed should be under 4000. The corresponding graphical formalism for this formula is presented in Fig. \ref{fig:safetyMultiple}.
\label{exmp:speedAndRpm}
\end{exmp}

The third type of template relationship enables the user to establish relative timing between two templates. Consider the following example:

\begin{exmp}

Specification $\phi_7 = \Box_{[0,40]}((speed < 80) \rightarrow \Box_{[0,40]} (rpm < 4000))$. Here, the nested specification $\Box_{[0,40]} (rpm < 4000)$ is evaluated every time $(speed < 80)$ is true. This formula is represented in the formalism with nested templates, otherwise referred to as parent and child templates. The second template is tabbed and connected to the first template using a green indicator. In the GUI, such a nested template is initiated by clicking on the signal of the parent template. The corresponding graphical formalism is presented in Fig. \ref{fig:nestedSequencing}.
\label{exmp:nestedSequence}
\end{exmp}

\begin{figure}[b]
\ifthenelse{\boolean{TECHREP}}{
  \vspace{10pt}
}{
  \vspacemod{-5pt} % in text
}

% \vspacemod{-10pt} % on top of page
\begin{center}
	\includegraphics[width=6.5cm]{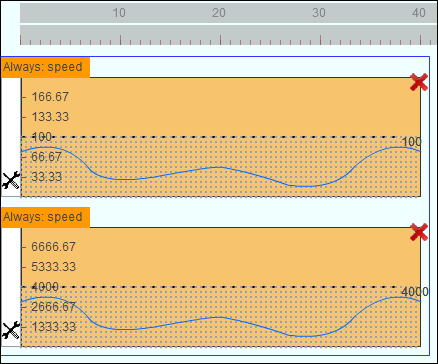}
\end{center}
\vspacemod{-5pt}
\caption{Example \ref{exmp:speedAndRpm}: The graphical formalism for the MTL specification $\phi_{6}=(\Box_{[0,40]}(speed < 100)) \wedge (\Box_{[0,40]} (rpm < 4000))$.}
\label{fig:safetyMultiple}
\vspacemod{-5pt} % in text
\end{figure}

\begin{figure}[thb]
\vspacemod{-5pt} % in text
% \vspacemod{-10pt} % on top of page
\begin{center}
\includegraphics[width=\columnwidth]{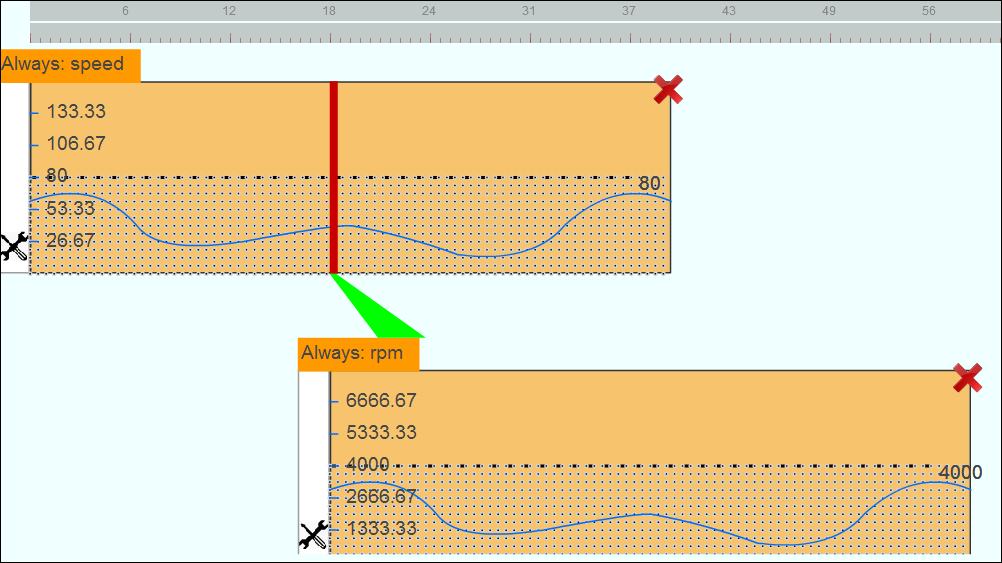}
\end{center}
\vspacemod{-5pt}
\caption{Example \ref{exmp:nestedSequence}: The graphical formalism for the MTL specification $\phi_7 = \Box_{[0,40]}((speed < 80) \rightarrow \Box_{[0,40]} (rpm < 4000))$.}
\label{fig:nestedSequencing}
\vspacemod{-10pt} % in text
\end{figure}
}
{
For more examples of specifications and their corresponding graphical formalism see the technical report \cite{Hoxha_ViSpecTechRpt15}. In Section \ref{sec:application}, we present two application specifications that illustrate the various interactions between templates.
}

The variety of templates and the connections between them allow users to express a wide variety of specifications.

\ifthenelse{\boolean{TECHREP}}{
  \begin{figure*}[tb]
\vspacemod{-5pt} % in text
% \vspacemod{-10pt} % on top of page
\begin{center}
\includegraphics[width=12.5cm]{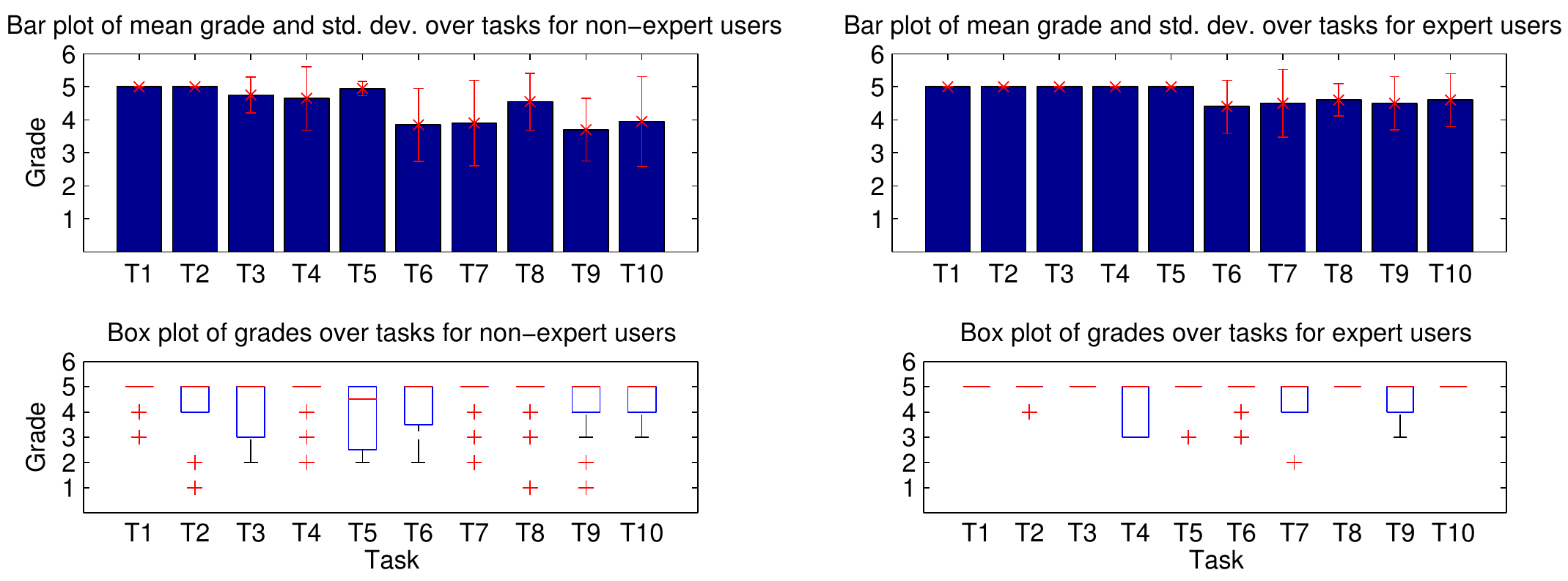}
\end{center}
\vspace{-5pt}
\caption{Subject accuracy grades over tasks for both the expert and non-expert cohorts.}
\label{fig:boxbarplotaveragegrade}
\vspace{-20pt} % in text
\end{figure*} 
}{}

%!TEX root = root.tex
\pdfoutput=1 
\section{Graphical Formalism}

\tikzstyle{int}=[draw, fill=blue!20, minimum size=2em]
\tikzstyle{init} = [pin edge={to-,thin,black}]

\begin{figure}[t]
\vspacemod{-0.1cm}
\centering
\begin{tikzpicture}[node distance=3.5cm,auto,>=latex']
    \node [int, text width=1.2cm,align=center] (a) {\ViSpec{} Tool};
    \node (b) [left of=a,node distance=2cm, coordinate] {a};
    \node [int] (c) [right of=a, text width=1.6cm,align=center] {Graphical Formalism};
    \node [coordinate] (end) [right of=c, node distance=2.8cm]{};
    \path[->] (b) [text width=1cm,align=center] edge node {User Input} (a);
    \path[->] (a) [text width=1.5cm,align=center] edge node {Tree Structure} (c);
    \draw[->] (c) [text width=0.8cm,align=center] edge node {MTL} (end) ;
\end{tikzpicture}
\caption{The specification development process using \ViSpec{}}
\label{fig:viSpecProcess}
\vspacemod{-0.02cm}
\end{figure}
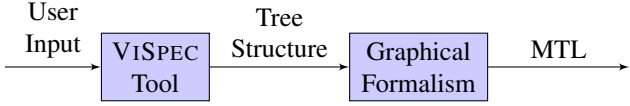

The specification development process in \ViSpec{} is divided in two sub processes. First, given a user input in the \ViSpec{} tool, it is translated to a tree structure where the nodes contain template information such as temporal operators, their corresponding timing parameters, group and the value threshold for the predicates. Secondly, the generated tree structure is traversed by a recursive algorithm to generate the MTL formula. There is a bijection between the visual representation of a specification and the MTL formula. An overview of the process is provided in Fig. \ref{fig:viSpecProcess}. 

An example of the tree structure for MTL formula $\phi = \Box(a \wedge \Diamond b) \rightarrow (\Box c \wedge \Diamond(d \rightarrow (a \wedge \Box b)))$ is shown in Fig. \ref{fig:treeStructure}. The recursive algorithm for traversing the tree structure and generating the MTL formula is presented in Alg. \ref{alg:writeMTL}. Note that the functions \text{\sc addParenConn\{A,B,C,D\}} add the parenthesis and connectives between predicates. 
\ifthenelse{\boolean{TECHREP}}
{}
{
A more detailed presentation of the algorithm is presented in the technical report \cite{Hoxha_ViSpecTechRpt15}. 
}

\tikzset{
  treenode/.style = {align=center, inner sep=3pt, text centered,
    font=\sffamily},
  arn_n/.style = {treenode, rectangle, black, draw=black,
    fill=white,  minimum width= 1cm},% arbre rouge noir, noeud noir
}

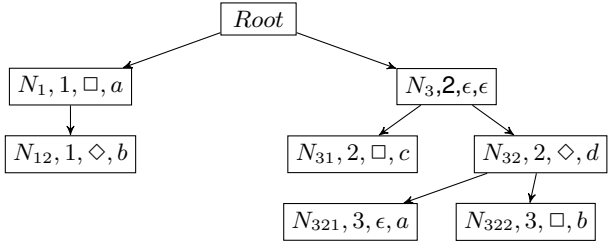
\begin{figure}[h]
\vspacemod{-0.2cm}
\centering
\begin{tikzpicture}[->,>=stealth',level/.style={sibling distance = 5cm/#1,
  level distance = 0.9cm}] 
\node [arn_n] {\small $Root$}
    child{ node [arn_n] {\small $N_1, 1, \Box, a$} 
            child{ node [arn_n] {\small $N_{12},1,\Diamond,b$}
            }                            
    }
    child{ node [arn_n] {\small $N_{3}$,2,$\epsilon$,$\epsilon$}
            child{ node [arn_n] {\small $N_{31},2,\Box,c$} 
            }
            child{ node [arn_n] {\small $N_{32},2,\Diamond,d$}
							child{ node [arn_n, left=0.8cm] {\small $N_{321},3,\epsilon,a$}}
							child{ node [arn_n, left=0.1cm] {\small $N_{322},3,\Box,b$}}
            }
		}
; 
\end{tikzpicture}
\caption{The corresponding tree structure for formula $\phi = \Box(a \wedge \Diamond b) \rightarrow (\Box c \wedge \Diamond(d \rightarrow (a \wedge \Box b)))$ where $a$,$b$,$c$ and $d$ are predicates. Each node is composed of a node name, group number, temporal operator, and predicate. The symbol $\epsilon$ indicates empty parameters. }
\vspacemod{-0.4cm}
\label{fig:treeStructure}
\end{figure}

\renewcommand{\algorithmicrequire}{\textbf{Input:}}
\renewcommand{\algorithmicensure}{\textbf{Output:}}

\ifthenelse{\boolean{TECHREP}}
{
\setlength{\textfloatsep}{0pt}
\begin{algorithm}
  \caption{WriteMTL - Algorithm for generating the MTL formula given a tree structure of the graphical formalism}
  \label{alg:writeMTL}
  \begin{algorithmic}[1]
    \Require Tree Structure $T=\tupleof{V,E}$ where $v \in V$ and $v = \tupleof{G,Op,S}$ where $G$ is the group, $Op$ is the temporal operator and $S$ is the predicate string; string $\phi$
    \Ensure $\phi$
    \Function {writeMTL}{$T,\phi$}
	\State $C$ $\gets$ $T.getChildren$. 
	\State $sC$ $\gets$ size($C$)
	\For {node $i$ in $C$}
		\State $\phi$ $\gets$ \text{\sc conc}($\phi$, $i.Op$)
		\If {$i$.isParent}
			\If {not($i.S.isEmpty$)}
				\State $subC$ $\gets$ $t.getChildren(i)$
				\If {$i.G == subC(1).G$}
					\State $\phi$ $\gets$ \text{\sc conc}($\phi$, '(', $i.S$, '$\wedge$')						
				\Else 
					\If {$i.isParent$}
						\State $\phi$ $\gets$ \text{\sc conc}($\phi$, '(', $i.S$,'$ \rightarrow ($')
					\Else
						\State $\phi$ $\gets$ \text{\sc conc}($\phi$, '(', $i.S$,'$ \rightarrow$')
					\EndIf
				\EndIf		
				\State $\phi$ $\gets$ \text{\sc writeMTL}($i.subtree$,$\phi$)
				\If {$i.isParent$}
					\If {$i.G == subC.G$}
						\State $\phi$ $\gets$ \text{\sc conc}($\phi$,'$)$')						
					\Else
						\State $\phi$ $\gets$ \text{\sc conc}($\phi$,'$))$')						
					\EndIf
				\Else
					\If {$sC$ $> 1$ and $i$ $\neq$ $sC$}
						\State $\phi$ $\gets$ \text{\sc conc}($\phi$,'$) \rightarrow$')	
					\Else
						\State $\phi$ $\gets$ \text{\sc conc}($\phi$,'$)$')
					\EndIf
				\EndIf				
			\Else 
				\State $\phi$ $\gets$ \text{\sc conc}($\phi$,'$($')
				\State $\phi$ $\gets$ \text{\sc writeMTL}($i.subtree$,$\phi$)					
				\If {i $\neq$ $sC$}
					\State $\phi$ $\gets$ \text{\sc conc}($\phi$,'$) \rightarrow$')	
				\Else
					\State $\phi$ $\gets$ \text{\sc conc}($\phi$,'$)$')
				\EndIf
			\EndIf
		\Else 
			\State $\phi$ $\gets$ \text{\sc conc}($\phi$,$i.S$)
			\If {$i$ $\neq$ $sC$}
				\State $\phi$ $\gets$ \text{\sc conc}($\phi$,'$\wedge$')	
			\Else
				\State $\phi$ $\gets$ \text{\sc conc}($\phi$,'$\rightarrow$')
			\EndIf							
		\EndIf
		
	\EndFor
	\EndFunction
  \end{algorithmic}
\end{algorithm}
}
{
\setlength{\textfloatsep}{0pt}
\begin{algorithm}
  \caption{WriteMTL - Algorithm for generating the MTL formula given a tree structure of the graphical formalism}
  \label{alg:writeMTL}
  \begin{algorithmic}[1]
    \Require Tree Structure $T=\tupleof{V,E}$ where $v \in V$ and $v = \tupleof{G,Op,S}$ where $G$ is the group, $Op$ is the temporal operator and $S$ is the predicate string; formula $\phi$.
    \Ensure $\phi$
    \Function {writeMTL}{$T,$ $\phi$}
	\State $C$ $\gets$ $T.getChildren$(). 
	\State $sC$ $\gets$ size($C$)
	\For {node $i$ in $C$}
		\State $\phi$ $\gets$ \text{\sc conc}($\phi$, $i.Op$)
		\If {$i$.isParent}
			\If {not($i.S.isEmpty$)}
				\State $subC$ $\gets$ $t.getChildren(i)$
				\State $\phi$ $\gets$ \text{\sc addParenConnA}($\phi$, $subC$)
				\State $\phi$ $\gets$ \text{\sc writeMTL}($i.subtree$, $\phi$)
				\State $\phi$ $\gets$ \text{\sc addParenConnB}($\phi$, $subC$)
			\Else 
				\State $\phi$ $\gets$ \text{\sc conc}($\phi$, '$($')
				\State $\phi$ $\gets$ \text{\sc writeMTL}($i.subtree$, $\phi$)		
				\State $\phi$ $\gets$ \text{\sc addParenConnC}($sC$, $\phi$)
			\EndIf
		\Else 
			\State $\phi$ $\gets$ \text{\sc conc}($\phi$, $i.S$)
			\State $\phi$ $\gets$ \text{\sc addParenConnD}($\phi$, $sC$)			
		\EndIf
		
	\EndFor
	\EndFunction
  \end{algorithmic}
\end{algorithm}
  
}

\vspacemod{-5pt}

\pdfoutput=1 
\section{Usability Study}

\subsection{Hypotheses}

The aim of the study is to evaluate whether \ViSpec{} enables users to develop formal specifications. Two groups were considered: 

\begin{enumerate}
\item Non-expert users: These are users who declared that they have no experience in working with requirements.
\item Expert users: These are users who declared that they have experience working with system requirements. Note that they do not necessarily have experience in writing requirements using formal logics. 
\end{enumerate}

Some of the interesting questions we wanted to investigate, which are also presented as hypotheses in Tab. \ref{tab:hypotheses}, are:

\begin{itemize}
\item Whether the graphical formalism enables non-experts and experts to formalize  requirements accurately.
\item How well the expert cohort performs in comparison to the non-expert cohort.
\item How user friendly and easy-to-use \ViSpec{} is. 
\end{itemize}

%%\normhyp
%\begin{subhyp}
%	\begin{hyp}
%	Non-expert users are able to define formal requirements accurately using formal logics such as MTL.
%	\label{hyp:nonExpMTL}
%	\end{hyp}
%	
%	\begin{hyp}
%	Non-expert users are able to define formal requirements accurately using the Visual Specification Tool.		\label{hyp:nonExpViSpec}
%	\end{hyp}
%\end{subhyp}
%\vspacemod{-0.6cm}
%
%\begin{subhyp}
%	\begin{hyp}
%	Expert users are able to define formal requirements accurately using formal logics such as MTL.
%	\label{hyp:ExpMTL}
%	\end{hyp}
%	
%	\begin{hyp}
%	Expert users are able to define formal requirements accurately using the Visual Specification Tool.
%	\label{hyp:ExpViSpec}
%	\end{hyp}
%\end{subhyp}
%
%Then the following natural hypothesis that arises:
%
%\noindent \textbf{Hypothesis $\mathbf{3_{null}}$} \textit{The average grade over all tasks for expert users is less than the average grade over all tasks for non-expert users.}
%
%\noindent \textbf{Hypothesis $\mathbf{3_{alt}}$} \textit{The average grade over all tasks for expert users is greater than or equal than the average grade over all tasks for non-expert users.}


\newcommand{\HypNonExpertMTL}{$1a$}
\newcommand{\HypNonExpertViSpec}{$1b$}
\newcommand{\HypExpertMTL}{$2a$}
\newcommand{\HypExpertViSpec}{$2b$}
\newcommand{\HypNonExpVExpNull}{$3_{null}$}
\newcommand{\HypNonExpVExpAlt}{$3_{alt}$}
\newcommand{\HypNonExpVExpTaskNull}{$Tx_{null}$}
\newcommand{\HypNonExpVExpTaskAlt}{$Tx_{alt}$}

\begin{table*}[htbp]
  \centering
  \caption{Hypotheses and test results with level of significance $\alpha = 0.05$. User groups as defined in section IV.A.}
    \begin{tabular}{p{0.5cm}p{13.3cm}c}
    \toprule
    Hypothesis & & Reject $null$ hypothesis \\
    \midrule
    \HypNonExpertMTL{}    & Non-expert users are able to define formal requirements accurately using formal logics such as MTL. &  \\
    \HypNonExpertViSpec{}    & Non-expert users are able to define formal requirements accurately using the Visual Specification Tool. & Yes \\
    \HypExpertMTL{}       & Expert users from the industry are able to define formal requirements accurately using formal logics such as MTL. &  \\
    \HypExpertViSpec{}    & Expert users from the industry are able to define formal requirements accurately using the Visual Specification Tool. & Yes \\
%    \HypNonExpVExpNull{} & The mean grade per user for expert users is less than or equal to the mean grade per user for non-expert users. & x \\
	\HypNonExpVExpAlt{} & The mean grade per user for expert users is greater the mean grade per user for non-expert users. & Yes \\
%    \HypNonExpVExpTaskNull{} & The mean grade per task x for industry users is less than or equal to the mean grade per task x for non-expert users.  & x \\
	\HypNonExpVExpTaskAlt{} & The mean grade per task x for industry users is greater than to the mean grade per task x for non-expert users.  & Partially \\
    \bottomrule
    \end{tabular}%
  \label{tab:hypotheses}%
  \vspacemod{-10pt}
\end{table*}%

Writing formal requirements is a challenging task that requires a significant amount of training. 
Therefore, it is safe to assume that we can reject Hypothesis \HypNonExpertMTL{} as supported by our informal experience. 
Hypothesis \HypExpertMTL{} will be tested in a future work.
% However, a related study with another formalism \cite{vinter1998applying} indicates that this can be an error prone process even for expert users.
In addition, we analyze user interaction and behavior to measure the ease-of-use of the tool.

%\begin{itemize}
%
%\item How do humans deal with the ambiguity? (Context-based disambiguation, Fluid meaning negotiation (dialectic) on the fly, )
%
%\item Hypotheses: 1- The Specification Visualization Tool will enable non-expert (no formal language experience) users to define formal requirements; 2- The Specification Visualization Tool will enable expert users to define formal requirements. 3-	The Specification Visualization Tool will enable users to develop requirements faster than Metric Temporal Logic 4-	The Specification Visualization Tool will enable users to capture requirements more accurately than Metric Temporal Logic
%
%\end{itemize}

\subsection{Demographics}

The non-expert cohort was comprised of twenty subjects from the student community of Arizona State University. Most of the subjects are from an engineering background with little to no experience working with requirements. The student demographics are presented in Tab. \ref{tab:demographics}.

The expert subject cohort was comprised of ten subjects from the industry in the Phoenix area. The subjects have experience working with specifications and come from an engineering background.

\begin{table}[htbp]
  %\vspacemod{-5pt}
  \centering
  \caption{Hypothesis \HypNonExpertViSpec{} Subject Demographics }
    \begin{tabular}{rrrrrr}
    Freshman & 2     & Computer Science & 5 & Male & 12 \\
    Sophomore & 2     &  Software Engineering & 3 & Female & 8  \\
    Junior & 5     & Electrical Engineering & 3 \\
    Senior & 5     & Mechanical Engineering & 6 \\
    Masters & 4     & Engineering, other & 3\\
    PhD   & 2     &       &  \\
    \end{tabular}%
  \label{tab:demographics}%
   \vspacemod{-1pt}
\end{table}%

\subsection{Experimental Design}

Each subject received a task list to complete. The task list contained ten tasks related to automotive system specifications. Each task asked the subject to formalize a natural language specification through \ViSpec{} and generate an MTL formula. The list of tasks is presented in Table \ref{tab:taskList}.

The tasks become more complex throughout the session. The higher the number of the task, the more steps necessary to complete the task successfully. 

Each session is at most 45 minutes long. Subjects received a one minute and thirty second tutorial on using \ViSpec{} to develop specifications. The computer screen was recorded and actions were logged for each session. The subjects also completed a demographic and post-completion questionnaire.

\subsection{Metrics}
%User performance evaluation is conducted with the following metrics:
Two metrics are used for performance evaluation:

\textit{Task completion}: this is a binary measure, which indicates whether users were able to finish the task within the set time. 

\ifthenelse{\boolean{TECHREP}}{
\begin{figure}[h]
%\vspacemod{-5pt} % in text
% \vspacemod{-10pt} % on top of page
\begin{center}
\includegraphics[width=6.5cm]{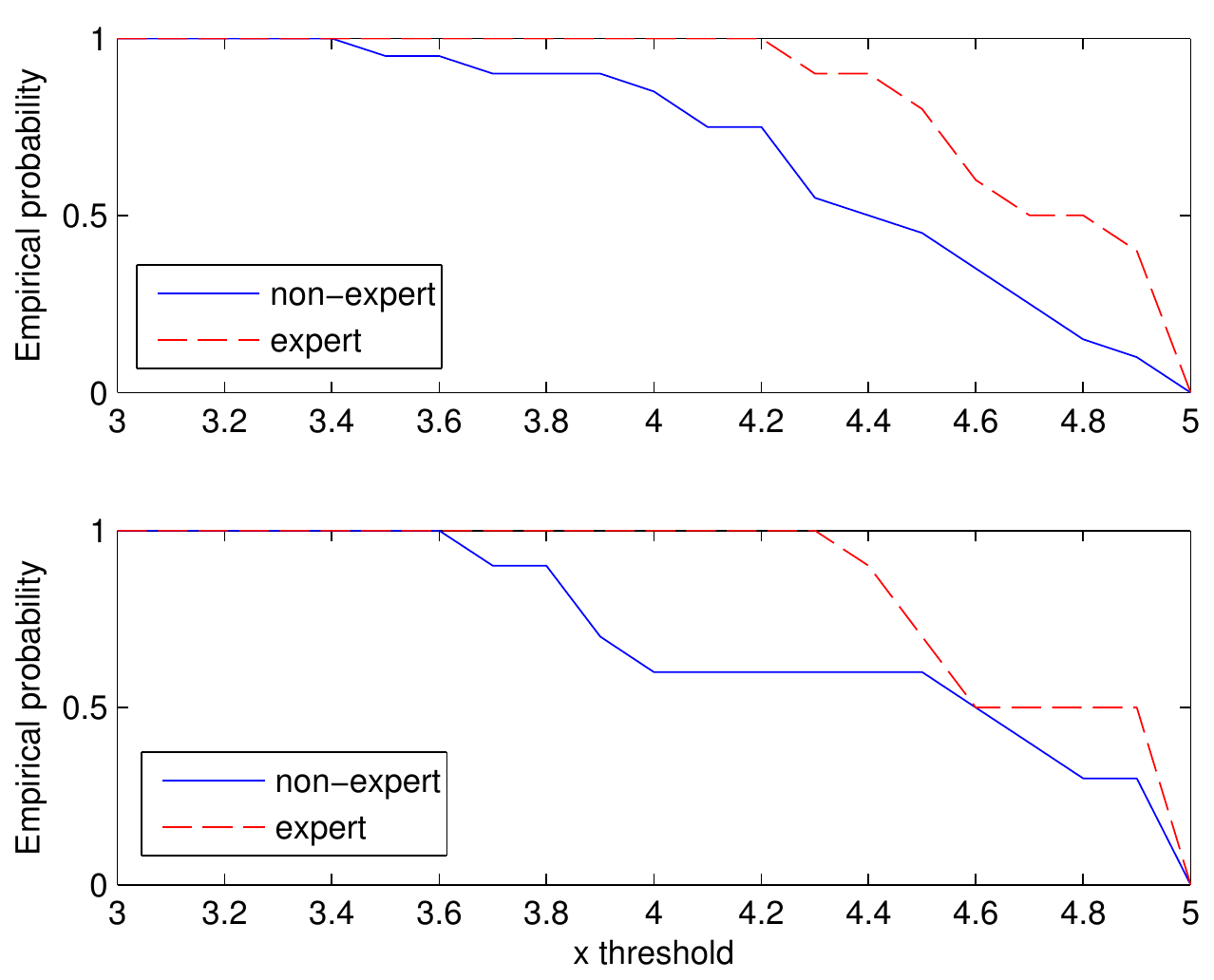}
\end{center}
\vspacemod{-5pt}
\caption{\textbf{Top}: The empirical probability that the mean grade per user is greater than threshold $x$ for the non-expert and expert subjects, i.e., $P(\bar{Y}>x)$. \\ \textbf{Bottom}: The empirical probability that the mean grade per task is greater than threshold $x$ for the non-expert and expert subjects, i.e., $P(\bar{X}>x)$. }
\label{fig:empiricalProbPerUserPerTask}
\vspacemod{-10pt} % in text
\end{figure}
}{}

\textit{Measure of Accuracy}: a value from one to five which is used to quantify the accuracy of subject generated formulas. The formulas are graded by formal specification experts which were given the following two suggested criteria: a) How accurate the meaning of the natural language specification is captured, and b) Whether the inaccuracies in the user submitted formula can be easily debugged and corrected in the testing and verification process. Furthermore, in order to decrease subjectivity, the following instructions were given to the expert graders in order to anchor the meanings of the five different grades of the scale used: A grade of one indicates that the generated formula is totally inaccurate. A grade of two indicates that the formula is mostly inaccurate. A grade of three indicates an inaccurate formula which can be easily debugged and corrected to the proper formal logic specification by formal specification experts and thus this is the minimum acceptable satisfactory result. A grade of four indicates that the formula is inaccurate but can be debugged and improved by automated specification debugging tools. A grade of five indicates that the generated formula is completely accurate. The group of expert graders consisted of experts in formal methods and logic.  

\section{Results}

\subsubsection{Average grade per task} 

For both cohorts, the task performance is presented in Fig. \ref{fig:boxbarplotaveragegrade}. It can be observed that overall, the mean grade per task for both cohorts is high. 
\ifthenelse{\boolean{TECHREP}}{Consider the mean grade per task as a random variable $\bar{X}$. Specifically, $\bar{X}:\Omega \rightarrow \Re$, where $\Omega \in \{y: 1 \leq y \leq 5 \}$.
In Figure \ref{fig:empiricalProbPerUserPerTask}, we present the survival function $S_{\bar{X}}(x) = 1 - F_{\bar{X}}(x) = 1 - P({\bar{X}}\leq x) = P({\bar{X}} > x)$ based on sample data. Note that $x$ is the threshold of mean grade accuracy. }{}

%To test Hypothesis \HypNonExpertViSpec, we would like to find the likelihood that a user achieves a grade higher than three when using the application, which we claim is an

\subsubsection{Hypothesis \HypNonExpertViSpec} 
To test Hypothesis \HypNonExpertViSpec, we need to establish what is an acceptable threshold for accuracy in order to test the hypothesis. As discussed in the metrics section, we claim that a mean grade higher than three is an acceptable threshold for non-expert users. Therefore, hypothesis \HypNonExpertViSpec{} is reduced to the null hypothesis: the mean grade per user is less than or equal to three for non-experts.

Let us define the average grade per user as a random variable $\bar{Y}$. Specifically, $\bar{Y}:\Omega \rightarrow \Re$, where $\Omega \in \{y: 1 \leq y \leq 5 \}$. The sample data from 20 subjects has a mean grade of 4.43 and standard deviation of 0.41. We test for normality with the Kolmogorov-Smirnov test, the Chi-square g.o.f test, and the Anderson-Darling test and all three fail to reject the null hypothesis that the data follows the normal distribution. \ifthenelse{\boolean{TECHREP}}{In figure \ref{fig:NormDataFit}, we plot the non-expert data against a fitted normal distribution and the corresponding Q-Q plot.}{} 
If we assume that the data constitute a random sample from a normal distribution, i.e. $\bar{Y} \sim \mathcal{N}$, we can use the t-statistic to test the hypothesis. We reject the null hypothesis with a p-value very close to 0. %of $1.38*10^{-12}$. 

\subsubsection{Hypothesis \HypExpertViSpec{}}
Similarly, we test Hypothesis \HypExpertViSpec{} for the expert cohort.
% However, the acceptable threshold for this cohort is to have a mean grade greater than 4. Here, Such specifications, if not completely accurate, can easily be debugged and possibly fixed by automated tools. 
Hypothesis \HypExpertViSpec{} is reduced to the null hypothesis: the mean grade per user is less than or equal to three for expert users. We test for normality as in the previous case and all three test fail to reject the null hypothesis that the data follows the normal distribution.  

Consider the average grade per user as a random variable $\bar{Z}$. Specifically, $\bar{Z}:\Omega \rightarrow \Re$, where $\Omega \in \{y: 1 \leq y \leq 5 \}$. The sample data from 10 subjects has a mean grade of 4.76 and standard deviation of 0.26. 
%We test for normality with the Kolmogorov-Smirnov test, the Chi-square g.o.f. test, and the Anderson-Darling test and all three fail to reject the null hypothesis that the data follows the normal distribution. 
\ifthenelse{\boolean{TECHREP}}{In figure \ref{fig:NormDataFit}, we plot the non-expert data against a fitted normal distribution and the corresponding Q-Q plot.}{} 
If we assume that the data constitute a random sample from a normal distribution, i.e. $\bar{Z} \sim \mathcal{N}$ we can use the t-statistic to test the hypothesis.We reject the null hypothesis with a p-value very close to $0$. 

\begin{figure}[h]
\vspacemod{-10pt} % in text
% \vspacemod{-10pt} % on top of page
\begin{center}
\includegraphics[width=7.5cm]{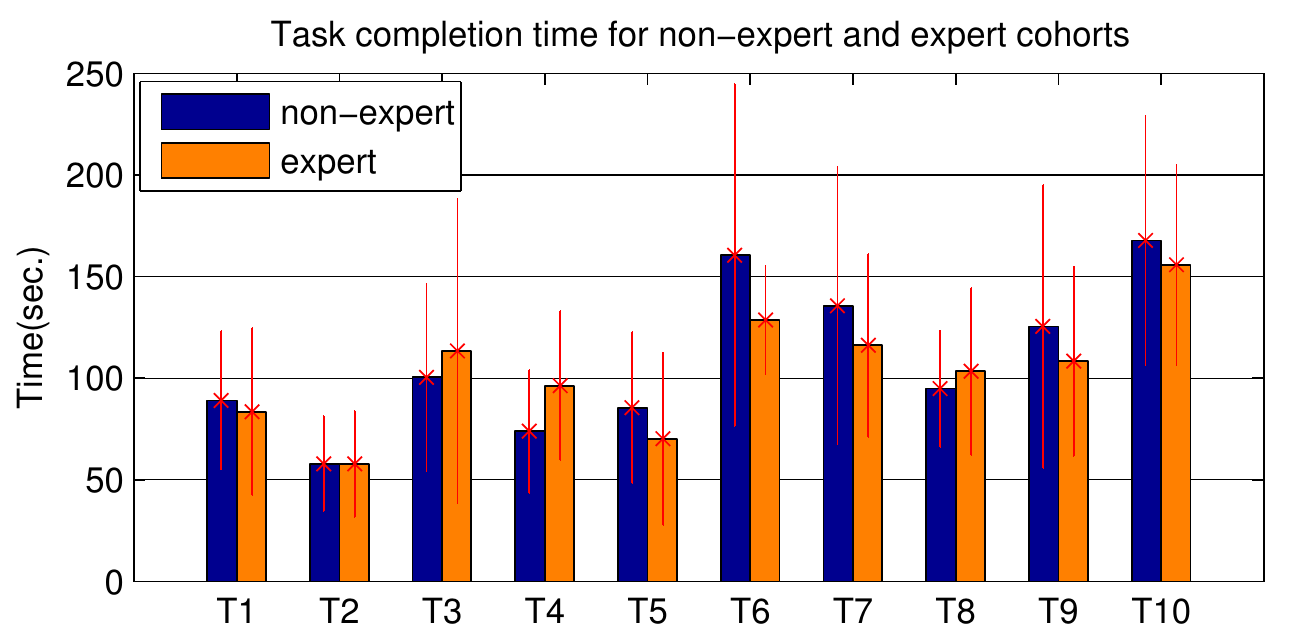}
\end{center}
\vspacemod{-5pt}
\caption{Example \ref{exmp:reachability}: The graphical formalism for the \textit{Reachability} MTL specification $\phi_{2}=\Diamond_{[0,39]} (speed > 100)$.}
\label{fig:taskCompletionTime}
\vspacemod{-15pt} % in text
\end{figure}

\begin{table}[htbp]
 \vspacemod{-2pt}
  \centering
  \caption{\ViSpec{} improvements}
    \begin{tabular}{p{0.1cm}p{4.4cm}p{2.8cm}}
    \toprule
    $\#$     & Improve... & Prime Indicators\\
    \midrule
    1.     & the process of creating child templates & misclicks; user feedback \\
    2.     & the tutorial by placing more emphasis on the difference between implication and conjunction between templates & task accuracy grade \\
    3.     & the visual representation of grouped templates & task accuracy grade; user feedback\\
  \bottomrule
    \end{tabular}%
  \label{tab:improvements}%
   %\vspacemod{-5pt}
\end{table}%

\subsubsection{Hypothesis \HypNonExpVExpAlt}

To test Hypothesis \HypNonExpVExpAlt, we conduct a two sample t-test. The p-value returned from the test is 0.0024 and for a significance level of 0.01, we reject the null hypothesis. Therefore we claim that the mean grade per user for expert users is greater than the mean grade per user for non-experts. 

\subsubsection{Hypothesis $Tx$}

Next, we compare the mean grade of both cohorts in regards to each task. A two sample \mbox{t-test} is conducted for each task. The results for the tests are presented in Tab. \ref{tab:TxHyp}. Task 9 is the most difficult task when it comes to the number of errors generated, and this is the only task where there is a clear difference in performance between the expert and non-expert cohorts. 

\begin{table}[htbp]
 %\vspacemod{-5pt}
  \centering
  \vspacemod{-5pt}
  \caption{Hypothesis testing of $Tx_{null}$ with $\alpha=0.05$}
    \begin{tabular}{rrrp{4.6cm}}
    \toprule
    $x$     & Rej. $Tx_{null}$ & p-val. & Conclusion \\
    \midrule
    4     & No     & 0.065  & potentially true with more investigation  \\
    5     & No     & 0.165  & false \\
    6     & No     & 0.074  & potentially true with more investigation \\
    7     & No     & 0.100  & potentially true with more investigation \\
    8     & No     & 0.424  & false \\
    9     & Yes    & 0.016 & true \\
    10    & No     & 0.063  & potentially true with more investigation \\
    \bottomrule
    \end{tabular}%
  \label{tab:TxHyp}%
   \vspacemod{-10pt}
\end{table}%

We observe that the only null hypothesis rejected is for task nine indicating that the mean grade for expert users is greater than the mean grade for non-expert users. The subject accuracy grades over tasks for is shown in Fig. \ref{fig:boxbarplotaveragegrade}.

\subsubsection{Ease-of-use analysis}

One indicator for the ease-of-use of the application is the total time spent per task. As can be observed in Fig. \ref{fig:taskCompletionTime}, the mean time spent per task on average is at most 167 seconds.  For easier identification of points of difficulty, we divided each task into subtasks. It was observed that there is no correlation between the length of time spent in a subtask and correctness.
%\ifthenelse{\boolean{TECHREP}}{(see Fig. \ref{fig:timeanalysis1} and \ref{fig:timeanalysis2})}{}. 
This potentially indicates, as also verified by correlation testing between times and grades, that the subjects were unaware of mistakes in the process. From these and other observations, such as misclicks, and subject feedback, we have developed a set of refinements on the tool to improve the user experience. A partial list of improvements is presented in Table \ref{tab:improvements}.

\begin{table*}[htbp]
  \centering
  \caption{Task list with automotive system specifications presented in natural language}
    \begin{tabular}{p{3cm}p{14cm}}
    \toprule
    Task  & Natural Language Specification \\
    \midrule
    1.   Safety  & In the first 40 seconds, vehicle speed should always be less than 160. \\
    2.   Reachability & In the first 30 seconds, vehicle speed should go over 120. \\
    3.   Stabilization & At some point in time in the first 30 seconds, vehicle speed will go over 100 and stay above for 20 seconds. \\
    4.   Recurrence & At every point in time in the first 40 seconds, vehicle speed will go over 100 in the next 10 seconds. \\
    5.   Recurrence & It is not the case that, for up to 40 seconds, the vehicle speed will go over 100 in every 10 second period.  \\
    6.   Implication & If, within 40 seconds, vehicle speed is above 100 then within 30 seconds from time 0, engine speed should be over 3000. \\
    7.   Reactive Response & If, at some point in time in the first 40 seconds, vehicle speed goes over 80 then from that point on, for the next 30 seconds, engine speed should be over 4000. \\
    8.   Conjunction & In the first 40 seconds, vehicle speed should be less than 100 and engine speed should be under 4000. \\
    9.   Non-strict sequencing & At some point in time in the first 40 seconds, vehicle speed should go over 80 and then from that point on, for the next 30 seconds, engine speed should be over 4000. \\
    10. Long sequence & If, at some point in time in the first 40 seconds, vehicle speed goes over 80 then from that point on, if within the next 20 seconds the engine speed goes over 4000, then, for the next 30 seconds, the vehicle speed should be over 100. \\
    \bottomrule
    \end{tabular}%
  \label{tab:taskList}%
   \vspacemod{-10pt}
\end{table*}%

\ifthenelse{\boolean{TECHREP}}{

\begin{figure*}[tbh]
\centering
\begin{subfigure}{.5\textwidth}
  \centering
  \includegraphics[width=6.5cm]{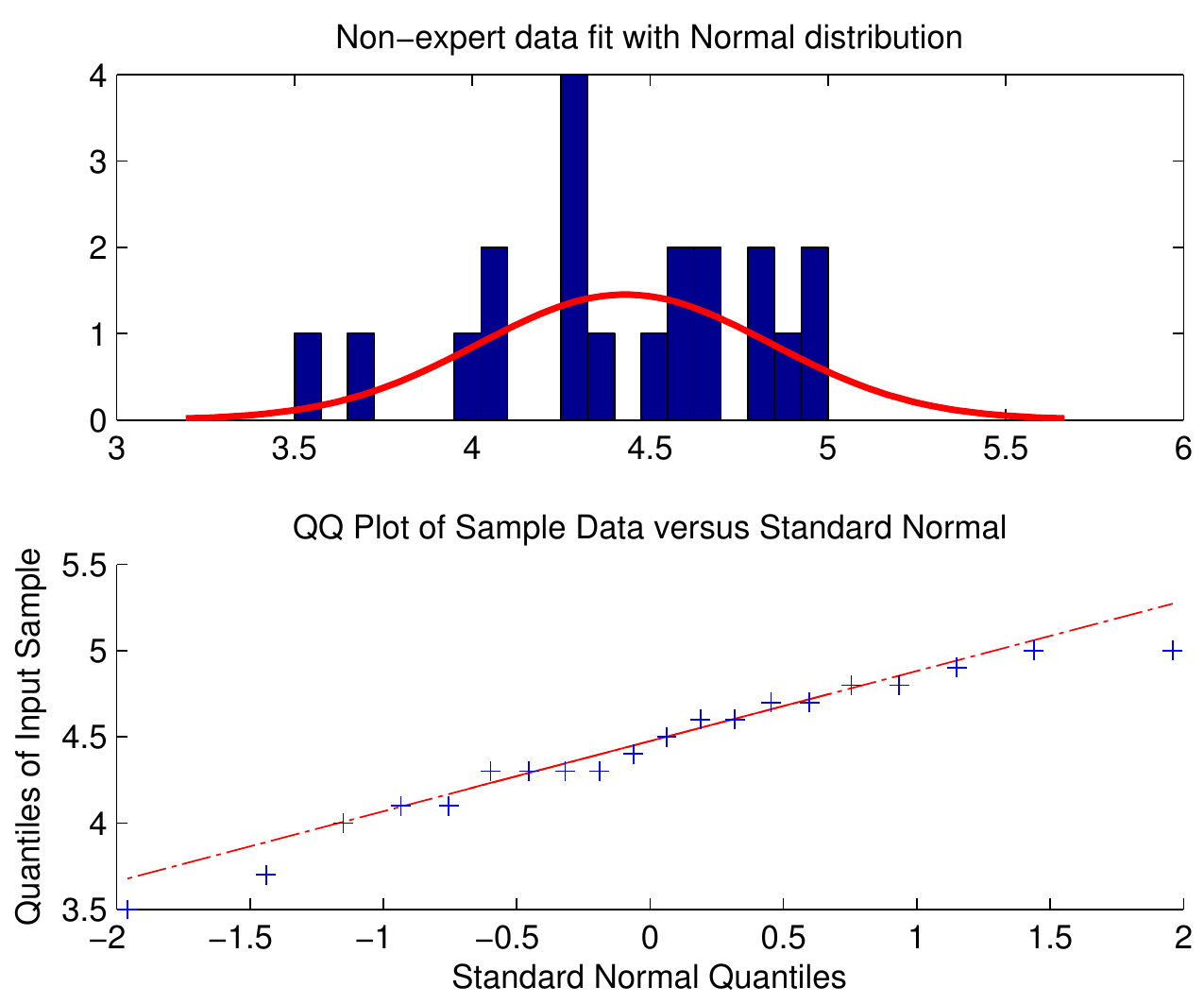}
  \label{fig:NonExpertNormDataFit}
\end{subfigure}%
\begin{subfigure}{.5\textwidth}
  \centering
  \includegraphics[width=6.5cm]{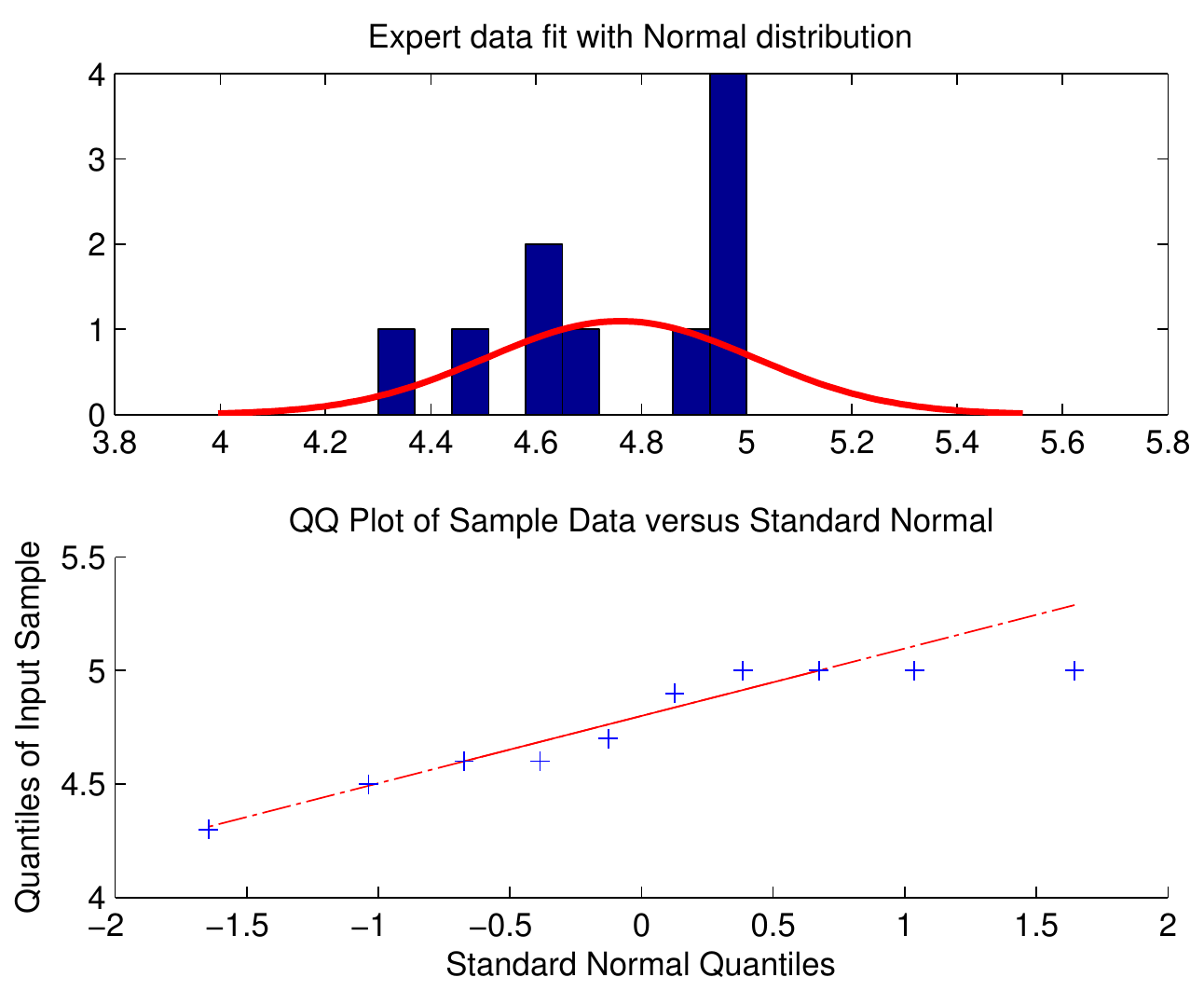}
  \label{fig:ExpertNormDataFit}
\end{subfigure}
\caption{Subject data fit with a normal distribution and the corresponding Q-Q plot.}
\label{fig:NormDataFit}
\vspace{-10pt}
\end{figure*}

}
{}

\ifthenelse{\boolean{TECHREP}}
{
%\afterpage{\clearpage}

%\begin{figure*}[tb]
%%\vspacemod{-5pt} % in text
%% \vspacemod{-10pt} % on top of page
%\begin{center}
%\includegraphics[width=\columnwidth+\columnwidth]{timeanalysis1}
%\end{center}
%%\vspacemod{-5pt}
%\caption{The subtask timing charts for tasks 1 through 4. The graphs represent the time spent on each part of the specification development process for both non-expert (blue) and expert (red) users.}
%\label{fig:timeanalysis1}
%\end{figure*}
%
%\afterpage{\clearpage}
%
%\begin{landscape}
% \begin{figure}[tb]
%  \centering
%  \includegraphics[width = 22cm]{timeanalysis2}
%  \caption{The subtask timing charts for tasks 6 through 10. The graphs represent the time spent on each part of the specification development process for both non-expert (blue) and expert (red) users.}
%  \label{fig:timeanalysis2}
% \end{figure}
%\end{landscape}
}{}

%\subsubsection{Average grade analysis} 
%
%The average grade, of all users, over all tasks is 

%\begin{itemize}
%\item Task completion: binary measure. Binary task completion includes measures of the number of correct tasks, the number of tasks where users failed to finish within a set time or the number of tasks where users gave up. In our case every student completed every task. 
%\item Measure of Completeness: Specification precision with grade (1-5). We grade the accuracy of formal requirements. 
%\item Indicators of thinking where users take long time between actions. 
%\item Indicators of confusion where users make a number of deletions.
%\item Measure periods of inactivity.
%\item Measure of accuracy: Number of errors in completing task. Number of attempts to complete task.
%\item Measure of efficiency: Time. Task completion time, Time in mode (The time users spend in a particular mode of interaction). Time until event (Time elapsed until users employ a specific feature or perform a particular action). 
%\item Usage patterns: Function frequency, deviation from optimal solution.
%\item Learning: users learning of the interface over tasks. 
%\item Measure of satisfaction: Overviews of expressed opinions about the interface. 
%\item How do we present all of this? 
%\end{itemize}

%!TEX root = root.tex
\pdfoutput=1 
\section{Applications}
\label{sec:application}

\subsection{Robotic Surgery} In the last few decades, there has been a significant increase in the number of robotics systems, especially in the health care system. They have been successfully introduced in multiple areas such as rehabilitation, telesurgery, physical therapy, elderly care, and remote physician care. In the following, we will focus on autonomous robotic systems for surgery where of paramount importance is the safety of these systems \cite{DBLP:conf/hybrid/KouskoulasRPK13}. Specifically, we will consider a model of a robotic serial link manipulator as presented in \cite{muradore2011robotic}.

One of the main tasks in surgery is the puncturing action. The high precision and repeatability of the process, make robot systems ideal for this task. Also, the trauma induced around the region is much lower and therefore the recovery process for the patient is quicker. To complete the puncturing action, the robot has to move towards the puncturing location. Test the tissue for various indicators to calibrate for optimal puncture, bring the puncturing needle to a perpendicular position and, finally, puncture with correct force and angle. If the force or angle is miscalculated, it might pose unintended harm to the patient. 
Consider the specifications from \cite{muradore2011robotic} that should hold on a serial manipulator for puncturing: 

\begin{enumerate}
\item From \cite{muradore2011robotic}: The force applied to the patient by the end effector is always less than a given threshold, except for the puncturing subtask. Formally, assuming that the operation time is 30 seconds, we have: $\phi_{s1} = \Box_{[0,30]}(\neg puncturing \rightarrow f \le f_{max})$.
\item From \cite{muradore2011robotic}: The task is feasible, and the position of the needle once it stops is inside the target region R. Formally, assuming that the operation time is 40 seconds, we have: $\phi_{s2} = \Diamond_{[0,40]}(Stop \wedge needle \in R))$.
\item Also, other requirements can be expressed for such a system. For example, the end effector speed should not be less than $v_{min}$ and should not be greater than $v_{max}$. Formally: $\phi_{s3} = \Box_{[0,40]}(v_{min} < v_{eff} < v_{max})$
\end{enumerate}

\begin{figure}[h]
\vspacemod{-5pt} % in text
% \vspacemod{-10pt} % on top of page
\begin{center}
\frame{\includegraphics[width=6cm]{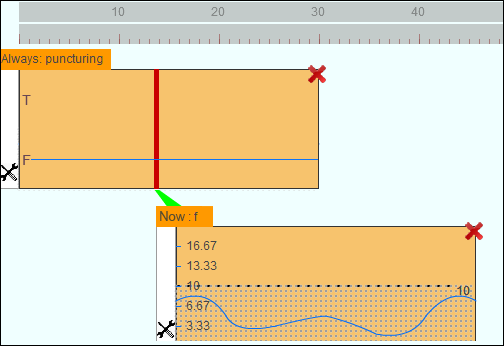}}
\end{center}
\vspacemod{-5pt}
\caption{The graphical formalism for $\phi_{s1}$.}
\label{fig:app_s1}
\vspacemod{-10pt} % in text
\end{figure}

The \ViSpec{} tool is utilized to develop the specifications for the robotic manipulator. For $\phi_{s1}$, the specification is presented in Fig. \ref{fig:app_s1}. We assume that $f_{max}=10$. 
\ifthenelse{\boolean{TECHREP}}
{For $\phi_{s2}$, the specification is presented in Fig. \ref{fig:app_s2}. We assume that $needle \in R \iff 5 < n_x < 10 \wedge 5<n_y <10$, where $n_x,n_y$ are the $x$ and $y$ coordinates for the needle.
For $\phi_{s3}$, the specification is presented in Fig. \ref{fig:app_s3}. We assume that $v_{min} = 10$ and $v_{max} = 20$. }
{
The visual representations with \ViSpec{} for item 2 and 3 are presented in the technical report \cite{Hoxha_ViSpecTechRpt15}. 
}

\subsection{Quadcopter} In recent years, quadcopters and other unmanned aerial vehicles (UAVs) have become a major focus for research both in the academic community and industry. Among others, they are used in military operations, nuclear disaster assessment, firefighting and entertainment. The challenges faced in developing these devices and their control algorithms come from the flight dynamics and the highly dynamical environment that they operate in. Also, as the complexity of these devices increases, so do the performance and reliability requirements.  

Consider the following specifications for a quadrotor:

\begin{enumerate}
\item The absolute value of the pitch and roll angle should always be bellow certain thresholds. 
Formally, assuming that the operation time is 40 seconds, we have: 
$\phi_{q1} = \Box_{[0,40]}(\left|\alpha\right| < \alpha_{max}) \wedge \Box_{[0,40]}(\left|\beta\right| < \beta_{max})$.
% $ \wedge \Box_{[0,40]}(\left|\gamma\right| < \gamma_{max})$. 
\item If distance to the target region is smaller than a certain threshold $d$, then for then next 20 seconds, the speed should not exceed $v_{max}$. Formally, assuming that the operation time is 40 seconds, we have: $\phi_{q1} = \Box_{[0,40]}(dist < d \rightarrow \Box_{[0,20]}(v < v_{max}))$. 
\end{enumerate}

\begin{figure}[h]
\vspacemod{-5pt} % in text
% \vspacemod{-10pt} % on top of page
\begin{center}
\frame{\includegraphics[width=6cm]{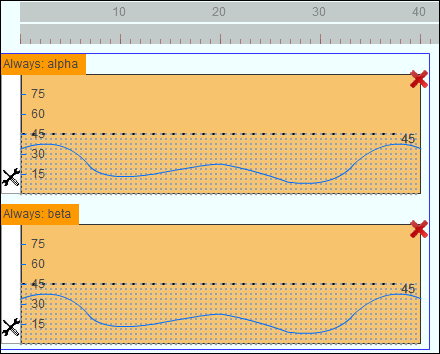}}
\end{center}
\vspacemod{-5pt}
\caption{The graphical formalism for $\phi_{q1}$.}
\label{fig:app_q1}
\vspacemod{-10pt} % in text
\end{figure}

The \ViSpec{} tool is utilized to develop the specifications for the quadrotor. For $\phi_{q1}$, the specification is presented in Fig. \ref{fig:app_q1}. We assume that $\alpha_{max} = 45\deg$, $\beta_{max} = 45\deg$ and $\gamma_{max} = 60\deg$. 
\ifthenelse{\boolean{TECHREP}}
{For $\phi_{q2}$, the specification is presented in Fig. \ref{fig:app_q2}. We assume that $d = 5 $ and $v_{max}=10$.
For $\phi_{s3}$, the specification is presented in Fig. \ref{fig:app_s3}. We assume that $v_{min} = 10$ and $v_{max} = 20$. }
{
The visual representations with \ViSpec{} for item 2 is presented in the technical report \cite{Hoxha_ViSpecTechRpt15}. 
}

\ifthenelse{\boolean{TECHREP}}
{
\begin{figure}[h]
\vspace{-5pt} % in text
% \vspacemod{-10pt} % on top of page
\begin{center}
\frame{\includegraphics[width=6cm]{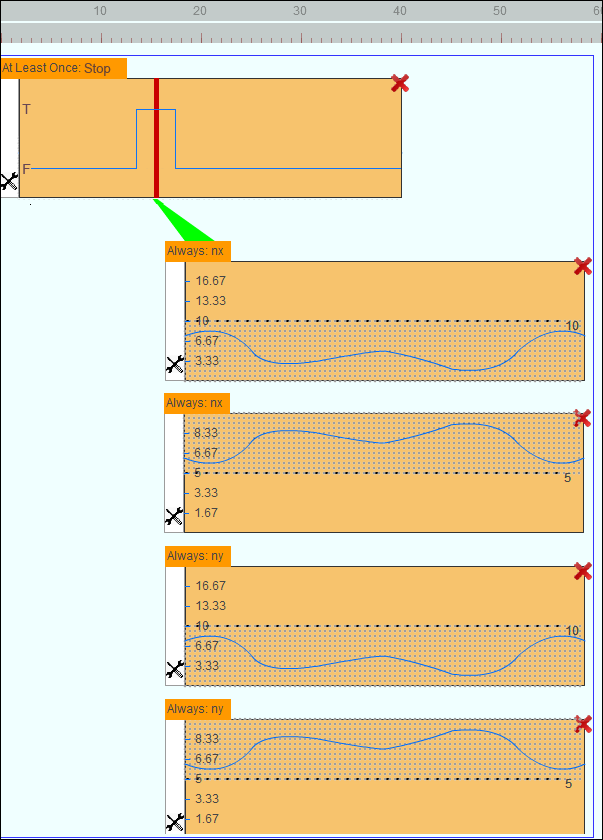}}
\end{center}
\vspace{-10pt}
\caption{The graphical formalism for $\phi_{s2}$.}
\label{fig:app_s2}
%\vspacemod{-10pt} % in text
\end{figure}

\begin{figure}[h]
\vspacemod{-15pt} % in text
% \vspacemod{-10pt} % on top of page
\begin{center}
\frame{\includegraphics[width=6cm]{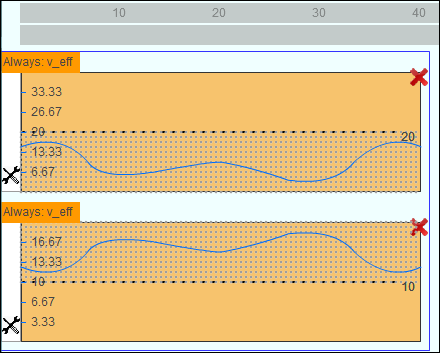}}
\end{center}
\vspace{-10pt}
\caption{The graphical formalism for $\phi_{s3}$.}
\label{fig:app_s3}
\vspacemod{-10pt} % in text
\end{figure}
}
{
}

\ifthenelse{\boolean{TECHREP}}
{

\begin{figure}[h]
\vspacemod{10pt} % in text
% \vspacemod{-10pt} % on top of page
\begin{center}
\frame{\includegraphics[width=6cm]{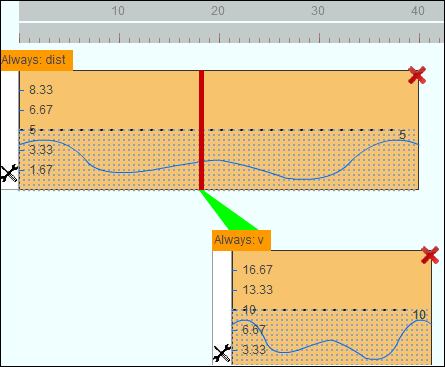}}
\end{center}
\vspace{-10pt}
\caption{The graphical formalism for $\phi_{q2}$.}
\label{fig:app_q2}
\vspacemod{-10pt} % in text
\end{figure}

%\begin{figure}[h]
%\vspacemod{-5pt} % in text
%% \vspacemod{-10pt} % on top of page
%\begin{center}
%\includegraphics[width=7.5cm]{quadrotor.jpg}
%\end{center}
%\vspacemod{-5pt}
%\caption{The AscTec Pelican Quadcopter.}
%\label{fig:quad}
%%\vspacemod{-10pt} % in text
%\end{figure}

}{}
\ifthenelse{\boolean{TECHREP}}{}{
\begin{figure*}[tb]
\vspacemod{-5pt} % in text
% \vspacemod{-10pt} % on top of page
\begin{center}
\includegraphics[width=12.5cm]{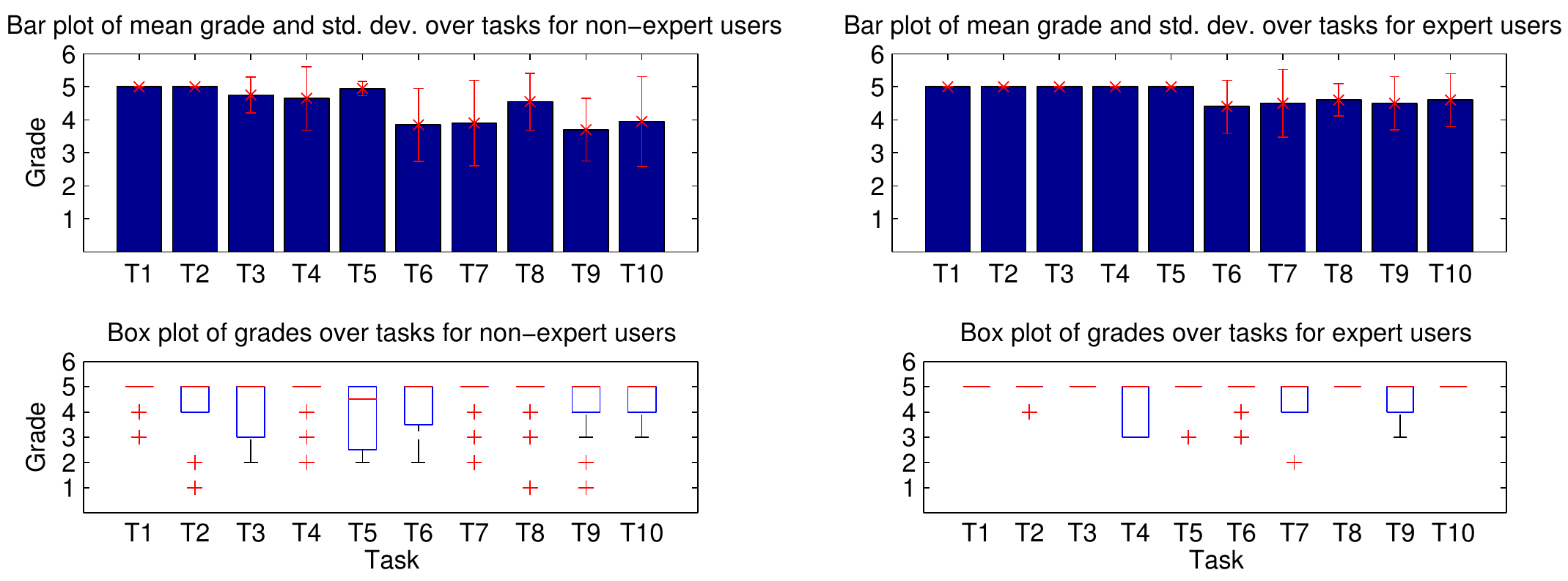}
\end{center}
\vspacemod{-5pt}
\caption{Subject accuracy grades over tasks for both the expert and non-expert cohorts.}
\label{fig:boxbarplotaveragegrade}
\vspacemod{-20pt} % in text
\end{figure*}
}
%\begin{figure}[h]
%\vspacemod{-5pt} % in text
%% \vspacemod{-10pt} % on top of page
%\begin{center}
%\includegraphics[width=\columnwidth]{application_item1.png}
%\end{center}
%\vspacemod{-5pt}
%\caption{The graphical formalism for the MTL specification $\Box(v_{eff} > 20) \wedge \Box(v_{eff} < 80))$.}
%\label{fig:item2}
%\vspacemod{-5pt} % in text
%\end{figure}

%!TEX root = root.tex
\pdfoutput=1 
\section{Conclusion and Future Work}
As robots and other cyber-physical systems become more complex and ubiquitous, so does the need for better testing and verification. A set of formal methods that improve this process require some formal representation of system specifications. In this work, a graphical formalism and a tool that enables users to easily develop formal specifications are presented. The \ViSpec{} tool enables users who have little to no mathematical training in formal logics to develop formal specifications, as was verified by a usability study that was conducted in order to evaluate
the usefulness of the tool and to get insights on potential improvements. The tool was utilized to formalize specifications for two robots.

Last but not least, we would like to investigate if the potential inaccuracies of the specifications that users generate with the tool can be attributed mainly to the inherent ambiguity of the natural language descriptions which were given, or if not, which other factors contribute and to what extent. Thus, in an improved usability study, we aim towards exploring alternative methods of generation of requirements from engineers for a system, that do not involve the administration of a natural language description by the experimenter. This would enable us to study to what extent inherent natural language ambiguity causes the observed less-than-perfect accuracy that is sometimes, even if rarely, exhibited.


%%%%%%%%%%%%%%%%%%%%%%%%%%%%%%%%%%%%%%%%%%%%%%%%%%%%%%%%%%%%%%%%%%%%%%%%%%%%%%%%

%%%%%%%%%%%%%%%%%%%%%%%%%%%%%%%%%%%%%%%%%%%%%%%%%%%%%%%%%%%%%%%%%%%%%%%%%%%%%%%%

%%%%%%%%%%%%%%%%%%%%%%%%%%%%%%%%%%%%%%%%%%%%%%%%%%%%%%%%%%%%%%%%%%%%%%%%%%%%%%%%
%\section*{APPENDIX}

\textsc{\textbf{Acknowledgment:}} Partial support under NSF awards CNS-1319560, CNS-1116136, IIP-1454143, IIP-1361926 and the NSF I/UCRC Center for Embedded Systems. We thank all the participants in the usability study and the reviewers for the detailed reviews. 

\bibliographystyle{abbrv}
\bibliography{fainekos_bibrefs,houssam,conformance,difts,conformanceMEMOCODE,bardh,adel}
%\nocite{*}
\end{document}